\documentclass[namedreferences]{solarphysics}
\usepackage[optionalrh,solaenum]{spr-sola-addons} 
\usepackage{graphicx}                    
\usepackage{color}                       
\usepackage{url}                         

\begin{document}

\begin{article}

\begin{opening}

\title{Turbulent convection in the Sun: modeling in unstructured meshes}

\author{V.~\surname{Olshevsky}$^{1}$\sep
        Ch.~\surname{Liang}$^{2}$\sep
        F.~\surname{Ham}$^{3}$      
       }

%
\runningauthor{Olshevsky et al.}
\runningtitle{Turbulent convection in the Sun: modeling in unstructured meshes}

%
\institute{$^{1}$ Main Astronomical Observatory, National Academy of Sciences, 27 Akademika Zabolotnoho st., 03680 Kyiv, Ukraine
                     email: \url{sya@mao.kiev.ua} \\ 
		  $^{2}$ Department of Mechanical and Aerospace Engineering, George Washington University, Washington, DC 20052 \\
              $^{3}$ Center for Turbulence Research, Stanford University, Stanford, CA 94305-3035                     
             }

%
\begin{abstract}
We adopted an unstructured hydrodynamical solver CharLES to the problem of global convection in the Sun.
With the aim to investigate the properties of solar turbulent convection and reproduce differential rotation pattern.
We performed simulations in two spherical shells, with 1.3 and 10 million cells.
In the first, coarse mesh, the solution does not reproduce realistic convection, and is dominated by numerical effects.
In the second mesh, thermal conduction leads to cooling of bottom layers, that could not be compensated by solar irradiance.
More simulations in the 10M cells mesh should be performed to investigate the influence of transport coefficients and numerical effects.
Our estimate of the code performance suggests, that realistic simulations in even finer grids could be performed for reasonable computational cost.
\end{abstract}

%

\end{opening}

\section{Introduction}\label{s:intro} 
This manuscript dates back to 2011, when we made first steps to model solar global convection using a high-end unstructured mesh CFD solver. Unfortunately, the paper was rejected from Solar Physics (as it described a work-in-progress), and we were unable to continue our studies. I hope that now, on the eve of the new 2015 year, and the era of exascale computing, this paper may inspire someone to create a new, fully compressible MHD model of the solar global convection. I am well aware of the new developments in the field, especially the exciting ``reduced sound speed'' technique, but I decided to keep this manuscript in its original, three-year-old, form. 

Stellar convection is a ``well known'', but poorly understood phenomena. 
The earliest attempts to study this process date back to the mid 20th century. 
They were motivated by stellar structure and evolution theory, which required the values of convective energy flux.
Stellar convection is a process where solution of hydrodynamic equations becomes very complicated.
Mostly because this process is highly turbulent. 
Due to low molecular viscosity, Reynolds number in the convective zone of the Sun reaches $10^{12}-10^{13}$. 
These values of $Re$ are far beyound reach of any earth-based laboratory experiment.
Picture of solar turbulent convection is further complicated due to rotation of the Sun.
As was pointed out by \opencite{thomp:etal:2003}, understanding the dynamics of such highly turbulent, complex structure as convective zone, must rely on numerical modeling. 

Unfortunately, computational capabilities were rather limited, and only last two decades gave rise 
to a number of realistic three-dimensional (3D) modeling attempts. 
Those can be splitted into two major groups: models of (sub-) surface convection, and models of the global-scale convection.
The former ones have reached great success in recent years, and agree remarkably well with observations \cite{lrsp-2009-2}\footnote{Here and below, if we refer to observations, we mean not only the data obtained from telescopes, but also the results of helioseismic inversions.}.
Unfortunately, global-scale simulations were not so successful, to our understanding.

So far, only a few numerical codes have been used for modeling stellar global-scale convection and differential rotation.
Majority of the simulations reported in the literature, were performed with Anelastic Spherical Harmonic (ASH) code developed by \opencite{clune:etal:1999}. 
ASH is a semi-spectral anelastic turbulent magneto-hydrodynamical (MHD) code, designed for spherical geometries. 
Its applications include simulations of solar convective zone \cite{miesch:2005,miesch:etal:2008}, 
tachocline studies \cite{brun:zahn:2006}, and dynamo simulations \cite{brown:etal:2010}. 
Recently, attempts have been made to model global-scale convection with Pencil code\footnote{\url{http://code.google.com/p/pencil-code/}}, 
developed at Nordita (Stockholm, Sweden).
Pencil is a Direct Numerical Simulation (DNS) MHD code designed for modeling weakly compressible turbulent flows, and is applicable to general astrophysical plasmas. 
Advances in simulations of large-scale stellar dynamos with Pencil were reviewed by \opencite{brand:2009}. 
A finite-volume hydrodynamical code EULAG \cite{prus:etal:2008}, has been adopted for global simulations of solar convection by \opencite{ghiz:etal:2010}.

Present-day theoretical and numerical models agree {\it qualitatively} with the observed characteristics of solar turbulent convection and differential rotation. 
To testify existing models, and provide capabilities for reaching {\it quantitative} agreement with observations, 
we have adopted a fully-compressible solver CharLES to the problem of solar global-scale convection.
With its capability to operate on unstructured grids, CharLES offers a promising opportunity 
to resolve the tachocline and subphotospheric shear layer together with convective zone in a single simulation domain. 
Because CharLES is a fully-compressible solver, it can deal with high entropy gradients, which is a problem for anelastic codes. 
The influence of different sub-grid scale (SGS) models on the simulations can be studied with CharLES.

\section{Method}\label{s:method}

The CharLES solver was developed at the Center for Turbulence Research of Stanford University \cite{ham08,ham10}.
It employs a hybrid second-order central-difference, finite-volume method and a WENO scheme to discretize the Navier-Stokes and conserved scalar equations on unstructured finite-volume grids. 
For the cell interface flux, an HLLC Riemann solver is employed. 
A three-stage explicit Runge-Kutta scheme is used for time marching. 
ParMetis \cite{kar:kum:1998} is employed for domain decomposition to parallel processors. 
Communication between the processors is handled using MPI.

\subsection{Equations in rotating frame}

To study stellar global-scale convection, the equations of hydrodynamics must be formulated in a rotating reference frame.
Consider a reference frame that is uniformly rotating with angular speed ${\mathbf\Omega}_0$. 
The velocity of the uniform rotation at $\mathbf{r}$ is $\mathbf{u}_0 = [\mathbf{\Omega}_0\times\mathbf{r}]$. 
In this noninertial frame, additional terms, representing centrifugal and Coriolis forces, should be included to the energy and momentum conservation laws. 
The resulting system of equations is

\begin{equation}
\frac{\partial \rho}{\partial t} + \nabla \cdot (\rho \mathbf{u}) = 0,
\label{eq:mass}
\end{equation}

\begin{equation}
{\partial(\rho \mathbf{u})\over\partial t}+ \nabla \cdot\left( \rho
\mathbf{u} \mathbf{u} + p \hat{\mathrm I}\right) = 
\nabla \cdot {\hat\tau} + \rho \mathbf{g} - 
\rho\left( 2\left[\mathbf{\Omega}_0\times \mathbf{u}\right] + 
\left[\mathbf{\Omega}_0\times \left[\mathbf{\Omega}_0\times\mathbf{r}\right]\right] \right),
\label{eq:threemomentum}
\end{equation}

\begin{equation}
{\partial\left(\rho E\right)\over\partial t} + \nabla\cdot\left( \left(\rho E+p\right)\mathbf{u}\right)= \nabla
\cdot \left( k\nabla T + \mathbf{u} \cdot \hat{\tau}\right) + \rho\mathbf{u}\cdot\mathbf{g} + Q_{\rm rad},
\label{eq:energy}
\end{equation}
where
\begin{equation}
\nonumber
  \rho E = {\epsilon} + {\rho u^{2}\over 2} - {\rho u_0^{2}\over 2}
\end{equation}
is the total energy per unit volume, $\epsilon$ is internal energy per unit volume, $\rho$ and $p$ are gas density and pressure, and $\mathbf{u}$ is velocity in the rotating frame. Symbol $\hat{\mathrm I}$ is a unit tensor, $\hat{\tau}$ is stress tensor, $\mathbf{g}$ is gravitational acceleration, $k$ is thermal conductivity coefficient, and $Q_{\rm rad}$ represents radiative source term. 
We introduced a tabulated OPAL equation of state\footnote{\url{http://opalopacity.llnl.gov/}} into the code to relate temperature and pressure with conserved quantities.

For a Newtonian fluid, the stress tensor is given by
\begin{equation}
\nonumber
\tau_{ij} = \mu\left[ {\partial u_i \over \partial x_j} + {\partial u_j \over \partial x_i} - {2 \over 3}\delta_{ij}\left(\nabla \mathbf{u}\right) \right],
\end{equation}
where $\mu = \rho\left(\nu_m + \nu_t\right) + \mu_l$ is the dynamic viscosity, $\nu_m$ is the molecular viscosity, and $\nu_t$ is the turbulent (eddy) viscosity, defined implicitly by the SGS model. 
Since the molecular viscosity $\nu_m$ in solar plasma is negligibly small, we drop the first term. 
The last term, $\mu_l$ is a predefined artificial dynamic viscosity, used to stabilize the solution. 

%
%
%
%
Similarly, the thermal conductivity, $k=C_p\rho\left(\kappa_m+\kappa_t\right) + k_l$, consists of three parts, associated with molecular diffusivity $\kappa_m$, eddy diffusivity $\kappa_t$, and an artificial conductivity $k_l$. 
Throughout the convective zone of the Sun, heat conduction is negligibly small compared with convective transport, and $\kappa_m$ may be ignored. 
In our simulations, the thermal conductivity coefficient is given by the relation: $k = C_p\mu/Pr$, where Prandtl number $Pr$ is kept constant throughout the computational domain.

The radiative source term is negligible compared to convective transport except in the uppermost part of the convective zone, $r>0.98R_\odot$ \cite{lrsp-2009-2}, and in the most of our simulation domain, we state $Q_{\rm rad}=0$.

\subsection{Boundary conditions}

Influence of boundary conditions on global convection simulations is rather poorly studied.
Some aspects of this problem were discussed by \cite{miesch:etal:2006}.
Most spherical shell simulations employ a constant energy inflow at the inner boundary, and either constant temperature, or constant energy outflow at the outer.
Use of such boundary conditions is possible only if anelastic or non-compressible fluid equations are solved.
In compressible flows, shocks are generated at the upper boundary, and upper layers are continuosly heated.
Following the strategy used in surface convection simulations, we introduce a thin layer with radiative cooling near the upper boundary.
In this layer, $Q_{\rm rad}$ term in Equation~(\ref{eq:energy}) is non-zero, and is given by Newton's cooling law:
\begin{equation}
  Q_{\rm rad} = - c_{\rm v} (T - T_{\rm ref})/\tau_r,
\end{equation}
where $c_{\rm v}$ is heat capacity at constant volume, $T_{\rm ref}$ is the reference temperature taken from solar structure model, $\tau_r$ is thermal relaxation time.
Due to low heat conductivity, $\tau_r$ in convective zone is rather high, but in the photosphere it is of order of seconds.
In our simulations, this term artificially reproduces the fact that convective energy transport becomes radiative in a very thin layer, and the energy is irradiated to the space. So we impose $\tau_r = 100$~s, slightly higher than photospheric values.

At the inner boundary, we implement a constant energy inflow that equals to solar irradiance.

\begin{figure}    
  \centerline{\includegraphics[width=0.99\textwidth,clip=]{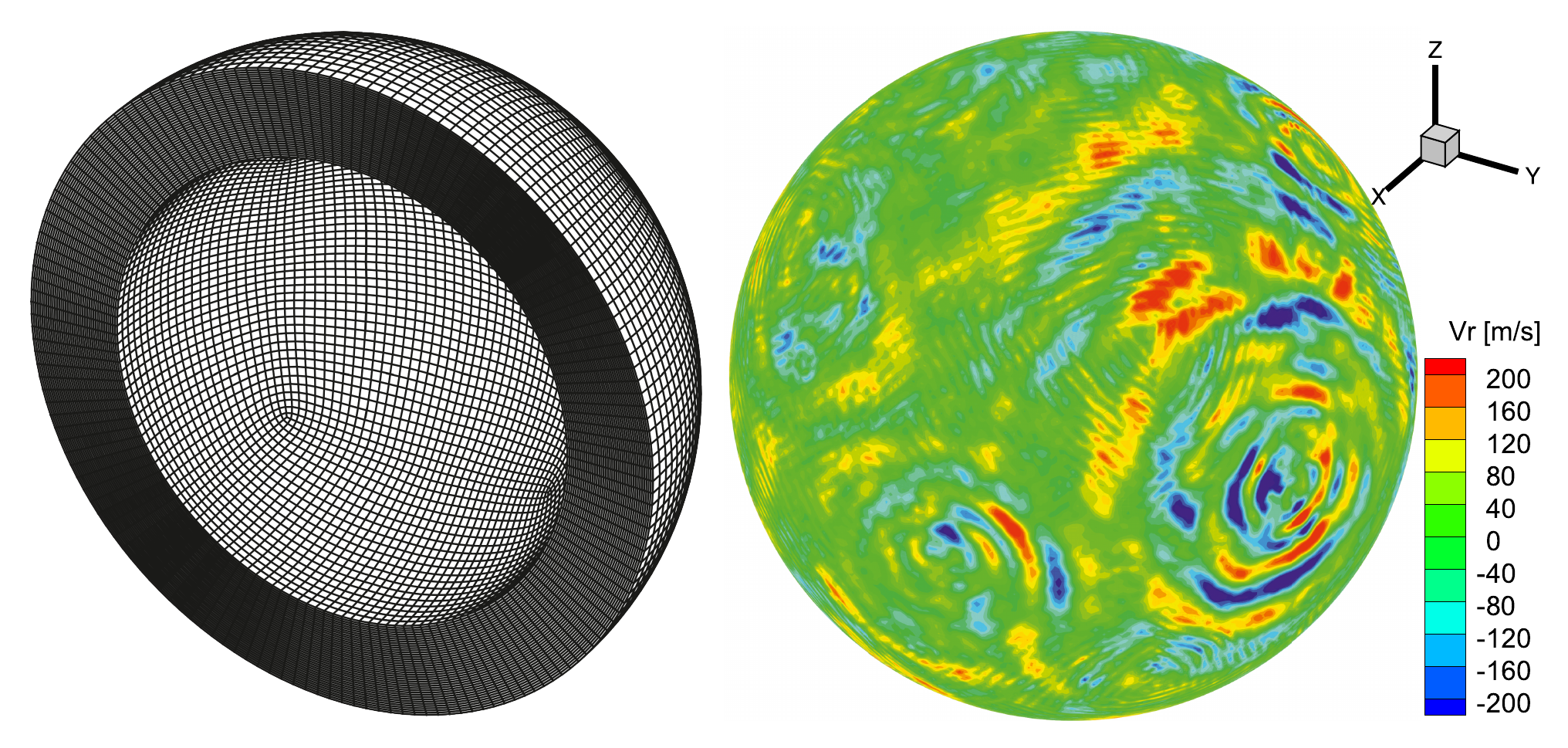}}
  \caption{Left: ``coarse'' mesh with 1.3M cells (only half of the cells are shown). Right: radial velocity at $r=0.82\,R_{\odot}$ on the 64-th day of solar time in the ``fine'' mesh simulation. The orientation of coordinate axes is the same in both panels.}
  \label{fig:1}
\end{figure}

\subsection{Mesh and simulation setup}

%
%
The simulations are performed in spherical shells that extend from the base of the convective 
zone, $r_{1}=0.7R_{\odot}$, to upper convective zone, $r_{2}=0.98R_{\odot}$.
In this work, two hexahedral grids generated by FLUENT and GAMBIT software\footnote{\url{http://www.ansys.com/}} were used. 
The first (coarse) mesh consists of 1.3M cells, while the second (fine) mesh has nearly 10M cells.
An example of computational mesh is shown in Figure~\ref{fig:1}. 
In the mesh, hexahedral cells are ordered in spherical layers.
On each spherical surface that contains cell vertices, each vertex joins 6 edges.
Except for 8 peculiar points, in which only 5 edges converge.

Initial unperturbed stratification of thermodynamic parameters depends only on radius $r$, and is taken from a standard solar structure model \cite{christ:etal:1996}.
The simulations are initialized ``as is'': from convectively unstable initial stratification, 
the instability develops rapidly, that first drives shocks, and then establishes convection.

\section{Results}\label{s:results}

\begin{table}
\caption{ Summary of simulation runs.
}
\label{tab:runs}
\begin{tabular}{rccccc}     
  \hline                   
No. cells & No. iterations & Solar time & Prandtl number & Viscosity & Comp. time\\
                &                   &       [days]     &  Pr                  &  $\mu_l$ [kg/m$\cdot$s]  &  [CPU hours]  \\
  \hline
$1\,296\,000$       & $1\,850\,000$  & 126  & 1.0 & $1.0\cdot 10^{10}$ & 30\,000 \\
$10\,368\,000$     & $2\,000\,000$  &   64  & 1.0 & $1.0\cdot 10^{10}$  & 290\,000 \\
  \hline
\end{tabular}
\end{table}

Solar energy flux is very small comparing to energy capacity in deep layers, $F_\odot / \rho c_s^3 \lesssim10^{-7}$ below $r\approx 0.95R_\odot$.
This flux is carried out by tiny temperature viriations $\sim 1$~K and at very small speeds $\lesssim 100$~m/s.
Therefore, the most complicated task, when using a low-order numerical scheme (and unstructured mesh), is to deal with such small fluctuations.
The runs reported here, were initialized without rotation induced, to investigate if the numerical precision allows to reproduce the correct convection picture.

We performed two simulations in different meshes, with otherwise identical parameters.
As mentioned above, the coarse mesh consists of 1.3M cells, and the fine one has 10M cells.
Table~\ref{tab:runs} lists the most important parameters of the two runs.
In both cases, Vreman SGS model was used \cite{vreman:2004} to compute turbulent viscosity.
The artificial viscosity $\mu_l$ was constant throughout the domain. 
We set it to the value two orders of magnitude smaller than in the simulations of \opencite{miesch:etal:2000}.

\begin{figure}    
  \centerline{\includegraphics[width=0.99\textwidth,clip=]{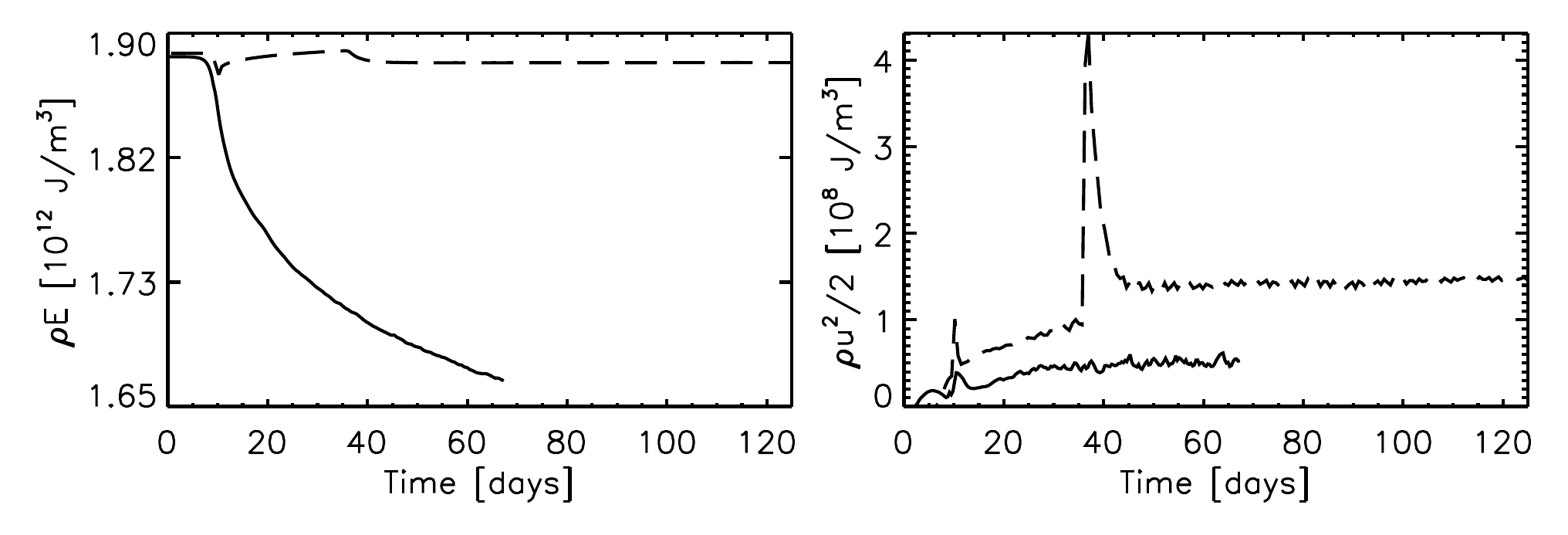}}
  \caption{Left: time dependence of the total energy density averaged over the simulation domain; Right: time dependence of the average kinetic energy density.
              Dashed lines correspond to the coarse mesh, and solid lines correspond to the fine mesh simulation.
              }
  \label{fig:2}
\end{figure}

In the coarse mesh, the simulation was running for nearly 126 days of solar time. 
Total and kinetic energy densities averaged over the simulation domains, are shown in Figure~\ref{fig:2}. 
The $\rho E$ varies slightly with time, decreasing from initial value by less than a percent, when the simulation reaches stationary state.
The kinetic energy grows as the instability develops. 
There are two prominent spikes in the $\rho u^2/2$ plot, at 10 and 40 days of solar time. 
The first one is associated with the interaction of shocks that develop in the unstable atmosphere at the very beginning of the simulation.
After the second spike, convection establishes in the coarse mesh simulation. 
After some 50 days of solar time, the fluctuations of $\rho u^2/2$ become very small, and we assume the stationary state has been reached in this simulation. 
However, the type of flow established, does not reproduce solar convection. 
The kinetic energy is about two orders of magnitude higher than is needed to carry solar energy flux. 
Consequently, a grid with 1.3M cells could not be used with CharLES solver to study solar convection.

\begin{figure}    
  \centerline{\includegraphics[width=0.99\textwidth,clip=]{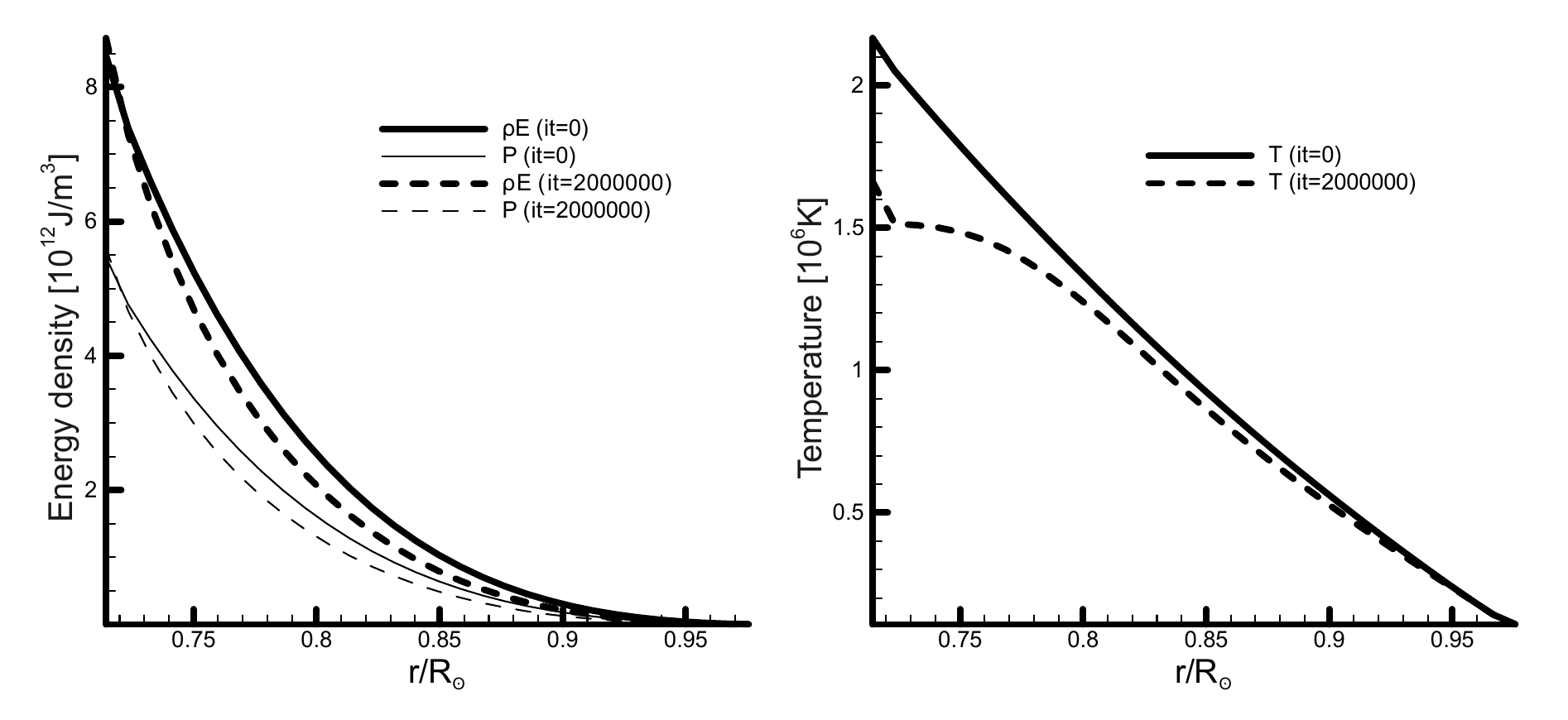}}
  \caption{Total energy density $\rho$E, pressure P and temperature T, averaged over longitude and latitude, for two time moments (the beginning and the end of the simulation) in the fine mesh.}
  \label{fig:3}
\end{figure}

The local maxima in $\rho u^2/2$ at 10th day of solar time is also present in the fine mesh simulation, but the second spike is absent.
The process of shock interaction is reproduced in both meshes, but in the fine mesh its energetics is much lower.
Although, the fluctuations of the average kinetic energy are small after some 50 days of solar time, the convection has not established in this case.
The flow is dominated by several huge vortices, noticeable in Figure~\ref{fig:1}. 
Those vortices are associated with peculiar grid points mentioned in Section~\ref{s:method}.
The values of $\rho u^2/2$ in the fine mesh are by order of magnitude smaller than in the coarse one.
Still, the velocities are too high to reproduce solar convection.
We decided to stop this simulation after only 64 days of solar time because of rapid decrease of the total energy density.
In Figure~\ref{fig:3} energy density, pressure and temperature is compared for the initial and the last time moments.
Due to thermal conductivity, the heat propagates up from the hottest layers at the base of convective zone.
In the Sun, the amount of energy going upward, is compensated by the radiative flux coming from the core.
In our simulations, the upward flux is too high, which quickly cools down the base, 
and leads to the creation of low-pressure zone in the middle of the shell ($r\approx 0.8R_{\odot}$).
Giant vortices transport matter downwards to compensate for cooling of the bottom layers.
Thus, more simulations in the fine mesh are needed to investigate if the cooling is caused by numerical effects,
or poorly chosen transport coefficients.

\section{Conclusions and discussion}\label{s:conclusions}

We have adopted an unstructured hydrodynamical solver CharLES to the problem of solar global convection.
We introduced the equations in a rotating reference frame, and an appropriate equation of state.
Constant energy inflow was induced on the bottom boundary, while on the upper boundary we use, an untypical
for such type of simulations, Newton cooling layer.

Two test runs were performed in different meshes: with 1.3M and 10M cells.
No rotation was induced, we only investigated the capability of the solver to reproduce solar convection.
The first mesh was clearly too coarse to model this process.
In the second mesh, the cooling of the bottom layers leaded to the creation of giant vortices in the middle of convection zone,
and no realistic convective pattern has been obtained.
More simulations in the 10M cells mesh are needed to investigate the influence of numerical inaccuracies and transport coefficients.

From our computations, the average performance of the code has been estimated.
It is about 45~$\mu$s per iteration per processor per cell in the coarse mesh, and slightly worse, 50~$\mu$s, in the fine mesh.
Probably, in the latter case the lower performance is due to bigger volumes of data output.
Also, the simulations in the fine mesh were running on 500---1000 processors, while in the coarse mesh only 96 processors were used.
With these, we can estimate that some 20M CPU hours are needed to simulate a year of solar time in a mesh with 100M cells.
Which is durable on the present-day supercomputers with a few thousands CPUs allocated for such simulation.

%
\begin{acks}
Authors acknowledge the support and useful discussions with Nagi Mansour and Alexander Kosovichev.

V. Olshevsky acknowledges the financial support to attend ESPM-13 provided by European Physical Society (EPS) Individual Member (IM) Travel Grant.

Authors acknowledge the ``MRI-R2: Acquisition of a Hybrid CPU/GPU and Visualization Cluster for Multidisciplinary Studies in Transport Physics with Uncertainty Quantification''\footnote{\url{http://www.nsf.gov/awardsearch/showAward.do?AwardNumber=0960306}} award for providing computing resources.
This award is funded under the American Recovery and Reinvestment Act of 2009 (Public Law 111-5).
\end{acks}

%
\bibliographystyle{spr-mp-sola}
 \bibliography{convection}

\end{article} 
\end{document}